\documentclass[11pt, a4paper]{article}
\usepackage{moriond,epsfig}

\def\be{\begin{equation}}
\def\ee{\end{equation}}
\def\bea{\begin{eqnarray}}
\def\eea{\end{eqnarray}}

\begin{document}
\vspace*{4cm}
\title{Charm and charmonium spectroscopy at B-factories}

\author{ Philippe Grenier (1,2)}
\address{(1) Stanford Linear Accelerator Center\\
2575 Sand Hill Road, MENLO PARK, CA-94025, USA}
\address{(2) Centre National de la Recherche Scientifique\\
3, Rue Michel Ange,  75794 Paris cedex 16 - France}

\address{Representing the Belle and Babar collaborations. Contributed to the proceedings of 
Moriond QCD 2007.}

\maketitle\abstracts{ We report on most recent Charm and Charmonium spectroscopy 
results from the B-factories.}

\section{Introduction}

Since a few years, charm and charmonium spectroscopy has revived, both from experimental and 
theoritical point of views. Many new states have been discovered triggering numerous theoritical 
publications. The B-factories with their large enriched charm sample have played a leading role 
on the experimental side with the observation and study of most of the new states. Other experiments 
such as CLEO and CDF  have also contributed. Classical hadron spectroscopy 
predicted some of these new states, but not all of them. Therefore a lot of effort have been spent 
in order to understand the nature of the later.  We are summarizing here the most recent and important 
results in hadron spectroscopy, including strange-charm mesons, charm baryons and charmonium 
and charmonium-like states.

\section{Strange-charm mesons $D_s$}

Prior to the B-factories, only 4 strange-strange mesons had been observed: the S-wave states 
$D_s(1968)^+$ ($J^P=0^-$) and $D_s(2112)^{*+}$ ($J^P=0^-$), and the P-wave states 
$D_{s1}(2536)^+$ ($J^P=1^+$) and $D_{s2}(2573)^{*+}$ ($J^P=2^+$). In 2003, the CLEO 
\cite{CLEODS2460} and Babar \cite{BABARDS2317} collaborations reported two states, 
the $D_{sJ}^*(2317)^+$ and the $D_{sJ}(2460)^+$, in continuum production ($e^-e^+ \to c \bar{c}$). 
These two states were subsequently observed by Belle \cite{BELLEDS} as wells as in B decays. The 
masses of the states were actually below expectations so there were a lot a speculation whether 
they were the missing $0^+$ and $1^+$ levels, other ($c \bar{s}$), or exotic states. The 
$D_{sJ}^*(2317)^+$ has been observed only in the $D_s^+ \pi^0$ decay mode, while the $D_{sJ}(2460)^+$ 
has been seen in the $D_s^{*+}(D_s^+ \gamma)\pi^0$, $D_s^+ \gamma$ and $D_s^+ \pi^+ \pi^-$ 
decay modes. Belle \cite{BELLEDS} and Babar \cite{BABARDSANG} have performed angular analysis of the $D_{sJ^*}(2317)^+ \to D_s^+ \pi^0$ and 
$D_{sJ}(2460)^+ \to D_s^+ \gamma$ decays and have shown that the two states are consistent with 
$J=0$ and $J=1$, respectively. A negative parity of the $D_{sJ}(2460)^+$ has been ruled out by an 
angular analysis of the $D_{sJ}(2460)^+ \to D_s^{*+}  \pi^0$ decay mode. Searches for isospin 
partners as well as for doubly charged states and different decay modes \cite{DS1} 
have been carried out. No signal has been found. All results indicate that the $D_{sJ}^*(2317)^+$ and the 
$D_{sJ}(2460)^+$ are indeed the missing $0^+$ and $1^+$ levels.

In 2006, Babar has reported the observation of a new $D_s$ meson \cite{BABAR2860} decaying into $D^0K^+$ and 
$D^+K_s$ with a mass $m=(2856.6\pm1.5\pm5.0) MeV/c^2$ and a width $\Gamma = (48\pm7\pm10) MeV$, 
where the first error is statistical and second is systematic. The new state has been observed in two 
inclusive continuum processes $e^+e^- \to D^0K^+X$ with $D^0 \to K^- \pi^+ , K^- \pi^+ \pi^0$ and 
$e^+e^- \to D^+K_SX$ with $D^+ \to K^- \pi^+ \pi^+$. The mass spectra of the three $D$ decay 
present similar features: a peak at $2.4 GeV/c^2$ due to a reflection of the $D_s(2536)^+$, 
a clear signal from the $D_s(2573)^+$, a broad structure peaking at a mass of approximately 
2700 $MeV/c^2$ and a signal enhancement due to the new state $D_s(2860)^+$. The broad 
structure at 2700 $MeV/c^2$ is fitted with a mass $m=(2688\pm4\pm3) MeV/c^2$ and a width 
$\Gamma = (112\pm7\pm36) MeV$. 

This later state was clearly observed in $B^+ \to \bar{D}^0 D_{sJ} \to  \bar{D}^0 D^0 K^+$ decay by 
the Belle collaboration \cite{BELLEDS2700}. It was then assigned to a new $D_{sJ}$ meson, the $D_{sJ}(2700)$. 
The Dalitz 
plot analysis shows that the decay $B^+ \to  \bar{D}^0 D^0 K^+$ proceeds dominantly through quasi-two-body 
channels: $B^+ \to \bar{D}^0 D_{sJ} (2700)^+$ and $B^+ \to \psi(3770) K^+$. The measured mass and width are 
$m=(2715\pm11^{+11}_{-14}) MeV/c^2$ and $\Gamma = (115\pm20^{+36}_{-32}) MeV$, respectively. The 
helicity angle distribution favors $J=1$. 

Babar has searched for resonances in $B \to \bar{D}^{(*)} D^{(*)} K$ decays, in 22 decay modes. 
The $DK$ and $D^*K$ invariant mass distributions have been built summing up all 8 corresponding B decay modes 
in each case. Both distributions present a clear enhancement near 2700 $MeV/c^2$. However, due to an 
unknown structure at low mass in the $DK$ invariant mass distribution and to the possible presence of 
additional resonances in the signal region in the $D^*K$ invariant mass distribution, no attempt has been 
made to extract the parameters of the the $D_{sJ}(2700)$. A full Dalitz plot analysis is undergoing.

\section{Charmonium(-like) states}

The X(3872) was discovered by Belle \cite{BELLE3872INIT} in $B^+ \to X(3872) K^+$ with 
$X(3872) \to J/\psi \pi^+ \pi^-$. It was confirmed \cite{BABAR3872} by CDF, D0 and 
Babar. There have been many speculations about the nature 
of this state: conventional charmonium state, charmonium hybrid, diquark-antidiquark state \cite{3872diquark}, 
or $D^0 \bar{D}^{*0}$ molecule \cite{molecule}. At present none of the hypothesis is favored. The current mass and 
width are $m=(3871.2\pm0.5) MeV/c^2$ and $\Gamma < 2.3 MeV$ (90$\%$ CL). Babar has found 
no evidence of a charged parnter. Belle \cite{BELLEXPIPI} has shown that 
the $\pi \pi$ invariant mass 
distribution 
favors positive parity $P=+1$. The CDF collaboration \cite{CDFXJPC} has performed an angular analysis of the 
$X(3872) \to \pi^+ \pi^-$ decay and demonstrated that quantum numbers $J^{PC}=1^{++}$ are 
favored. In parallel, Belle \cite{BELLEXGAM} and Babar \cite{BABARXGAM} have observed the decay 
$X(3872) \to J/\psi \gamma$ 
which indicates that the charge conjugation number is $C=+1$. 

Recently, Belle \cite{BELLEXDD} has observed an enhancement in the $D^0 \bar{D}^0 \pi^0$ system from 
$B \to D^0 \bar{D}^0 \pi^0 K$ decay. The enhancement peaks at a mass $m=(3875.4 \pm 0.7 ^{+1.2}_{-2.0})MeV/c^2$. 
Babar confirmed this observation in the following decays: 
$B^{+/0} \to \bar{D}^0 D^{*0} K^{+/0} \rm{and} B^{+/0} \to \bar{D}^{*0} D^{0} K^{+/0}$ with 
$D^{*0} \to D^0 \pi^0$ and $D^{*0} \to D^0 \gamma$. The ratio of neutral to charged modes branching 
fractions is $R = 2.23 \pm 0.93 \pm 0.55$. The combined mass from all 4 modes is 
$m=(3875.6\pm0.7^{+1.4}_{-1.5})MeV/c^2$. 
Therefore, the mass measurements from Belle and Babar are in very good agreement. However, it is more than 
2.5 standard deviations above the mass of the X(3872). Are they the same states or is the state at 3875 $MeV/c^2$ 
a new resonance?

Babar \cite{BABAR4260} has recently discovered a new state, the Y(4260) decaying in to $J/\psi \pi^+ \pi^-$, 
in the initial state radiation process $e^+ e^- \to \gamma_{ISR} J/\psi \pi^+ \pi^-$. The measured mass and 
width are $m=(4259\pm8\pm4) MeV/c^2$ and $\Gamma = (88\pm23\pm5) MeV$, respectively. 
The quantum numbers are straigthforward: $J^{PC}=1^{--}$. This state is still the subject of 
many theory papers attempting to explain its nature: classical charmonium state, tetraquark \cite{Ytetraq}, 
or hybrid charmonium \cite{Yhybrid1,Yhybrid2,Yhybrid3}. The Y(4260) was subsequently observed by the 
Cleo-c \cite{CLEOCY},  Cleo-III \cite{CLEOIIIY} 
and Belle collaborations \cite{BELLEY}. Belle's mass measurement is higher compared to Babar: 
$m=(4295\pm10^{+11}_{-5}) MeV/c^2$ with $\Gamma = (133^{+26+13}_{-22-6}) MeV$. 

Babar has been searching for $Y(4260) \to \psi(2S) \pi^+ \pi^-$ decay \cite{BABARY4325}. The observed 
$\psi(2S) \pi \pi$ invariant mass distribution shows an enhancement that is not compatible with the Y(4260), 
but with a higher mass resonance. Assuming the enhancement is due to a single resonance, the parameters of 
this resonance  would be $m=(4324\pm24) MeV/c^2$ and $\Gamma = (172\pm33)MeV$ where the errors are statistical 
only. We shall note that this enhancement is compatible with the Y(4260) from Belle which measures a higher 
mass. 

Finally, Belle has been reporting three states near $3940 MeV/c^2$. The Z(3930) has been observed 
\cite{BELLEZ3930} in $\gamma \gamma \to D \bar{D}$, with a mass $m=(3929\pm5\pm2)MeV/c^2$ and a 
width $\Gamma=(29\pm10\pm2) MeV$. The angular distribution of the decay strongly favors $J=2$. This 
state is interpreted as the ($c\bar{c}$) $2^3P_2 (2^{++})$ state (the $\chi^\prime_{c2}$). The Y(3940) 
has been observed \cite{BELLEY3940} in B decays $B \to J/\psi \omega(\pi \pi \pi)K$, with a mass 
$m=(3943\pm11\pm13)MeV/c^2$ and a width $\Gamma=(87\pm22\pm26) MeV$. This state has been tentatively 
assigned to the  ($c\bar{c}$) $2^3P_1 (1^{++})$ state (the $\chi^\prime_{c1}$) \cite{GODFREY}. The 
X(3940) has been observed \cite{BELLEX3940} in continuum production, in the recoil of a $J/\psi$: 
$e^+e^- \to J/\psi X$, with a mass $m=(3943 \pm 6 \pm 6)MeV/c^2$ and a width $\Gamma=(15.4\pm10.1) MeV$, 
and in the $X \to D \bar{D}^*$ decay mode. This state is likely to be the  ($c\bar{c}$) $3^1S_0 (1^{++})$ 
state (the $\eta_c(3S)$) \cite{GODFREY}.

\section{Charm baryon}

With the discovery of the $\Omega_c^*$ \cite{BABAROMEGACS}, all nine $J^P={1 \over 2}^+$ and six 
$J^P={3 \over 2}^+$ ground states ($L=0$) have now been observed. Several orbitally excited states have 
already been seen as well.

Babar \cite{BABAR2940} has observed a new state, the $\Lambda_c(2940)^+$, decaying into $D^0p$, with a 
mass $m=(2939.8\pm1.3\pm1.0)MeV/c^2$ and a width $\Gamma=17.5\pm5.2\pm5.9 MeV$. In this decay mode, 
Babar confirms the previously reported \cite{CLEO2880}  $\Lambda_c(2880)^+$ 
(seen in $\Lambda_c(2880)^+ \to \Lambda_c^+ \pi^+ \pi^-$), with 
a mass $m=(2881.9\pm0.1\pm0.5)MeV/c^2$ and a width $\Gamma=(5.8\pm1.5\pm1.1) MeV$.  The two states have 
also been reported by Belle \cite{BELLE28802940}, in $\Lambda_c^+ \pi^+ \pi^-$, with masses and widths 
$m=(2881.2\pm0.2^{+0.4}_{-0.3})MeV/c^2$ and $\Gamma=(5.5^{+0.7}_{-0.3}\pm0.4) MeV$, and 
$m=(2937.9\pm1.0^{+1.8}_{-0.4})MeV/c^2$ and a width $\Gamma=(10\pm4\pm5) MeV$, for the $\Lambda_c(2880)+$ 
and $\Lambda_c(2940)+$ respectively. The results from Belle and Babar are in very good agreement. Belle 
has also performed an angular analysis of the $\Lambda_c(2880)^+ \to \Sigma_c(2455)^{++/0}\pi^{0/+}$ and 
has shown that $J \ge 5/2$ is favored. 

Belle \cite{BELLE29803077} has observed two new $\Xi_c$ states decaying into $\Lambda_c^+ \pi^- \pi^+$, 
the $\Xi_c(2980)^+$ with $m=(2978.5\pm2.1\pm2.0)MeV/c^2$ and $\Gamma=(43.5\pm7.5\pm7.0) MeV$, 
and the $\Xi_c(3077)^+$ with $m=(3076.7\pm0.9\pm0.5)MeV/c^2$ and $\Gamma=(6.2\pm1.2\pm0.8) MeV$. 
Babar \cite{BABAR29083077} has confirmed these observations and measured the following parameters 
$m=(2967.1\pm1.9\pm1.0)MeV/c^2$ and $\Gamma=(23.6\pm2.8\pm1.3) MeV$ and 
$m=(3076.41\pm0.69\pm0.21)MeV/c^2$ and $\Gamma=(6.2\pm1.6\pm0.5) MeV$, respectively for the $\Xi_c(2980)^+$ 
and the $\Xi_c(3077)^+$.

The electromagnetic decay $\Xi_c^\prime \to \Xi_c \gamma$ has been measured by 
Babar \cite{BABARCASCADEP}. It is a confirmation of the Cleo observation. Contributions 
from B decays and from continuum are separated with the use of a cut on the $\Xi_c$ momentum 
measured in the center of mass frame. Branching fractions measured from production in B decays are 
$B(B \to \Xi_c^{\prime+}X) \times B(\Xi_c^+ \to \Xi^- \pi^+ \pi^-)= (1.69\pm0.17\pm0.10)10^{-4}$ 
and $B(B \to \Xi_c^{\prime 0}) \times B(\Xi_c^0 \to \Xi^- \pi^+ )= (0.67\pm0.07\pm0.03)10^{-4}$. 
For production from the continuum the cross sections are found to be 
$\sigma(e^+e^- \to \Xi_c^{\prime +}X) \times B(\Xi_c^+ \to \Xi^- \pi^+ \pi^+)=141\pm24(exp)\pm19(model) 
\rm{fb}$ 
and 
$\sigma(e^+e^- \to \Xi_c^{\prime 0}X) \times B(\Xi_c^0 \to \Xi^- \pi^+)=70\pm11(exp)\pm6(model) \rm{fb}$. 
The helicity angle distributions of $\Xi_c^{\prime}$ decays are found to be consistent with $J={1 \over 2}$.

Using a large $\Lambda_c$ sample, Babar \cite{BABARLAMBAC} has studied $\Lambda_c \bar{\Lambda}_c$ correlation 
production $e^+e^- \to \Lambda_c \bar{\Lambda}_c X$, where the $\Lambda_c \to $ is reconstructed in the 
$pK^-\pi+$ and $pK^s$ decay modes and the $\bar{\Lambda}_c$ in the corresponding charge-conjugate 
modes. The number of observed events is roughly 4.2 times the number of expected events with 
respect to models with at leasts four baryons in the final state. These events show very few additional 
baryons but multiple mesons, indicating a previously unobserved type of $e^+e^- \to q \bar{q}$ events.

\section*{Conclusion}
There have been a lot of activity in charm and charmonium spectroscopy at the B-factories 
during the last few years. Some of the newly discovered states match theoritical expectations, 
but most of them, such as the X(3872) and the Y(4260) are still to be understood.

\end{document}